\documentclass[aps,prd,reprint]{revtex4-2}

\usepackage{xspace}
\usepackage{lipsum}
\usepackage{relsize}
\usepackage{appendix}
\usepackage{amsmath,amstext,amssymb,amsfonts}
\usepackage{epsfig}
\usepackage{url}
\usepackage{nameref}
\usepackage{varioref}
\usepackage{hyperref}
\usepackage{cleveref}
\usepackage[normalem]{ulem} 

\usepackage{latexsym}
\usepackage{epsfig}
\usepackage{wasysym}
\usepackage{graphicx}
\usepackage{verbatim}
\usepackage{tikz} %diagrams
\usepackage{pgfplots}
\usepackage[export]{adjustbox}

\usepackage{bbold}

\usepackage[normalem]{ulem}%strikethrough text

\newcommand{\CMU}{McWilliams Center for Cosmology, Department of Physics, Carnegie Mellon University, Pittsburgh, PA 15213, USA}

\newcommand{\plpeak}{\textsc{PLB}\xspace}

\newcommand{\plplpeak}{\textsc{PLPLB}\xspace}

\graphicspath{{./}{figures/}}

\begin{document}

\title{Astrophysics informed Gaussian processes for gravitational-wave populations: Evidence for the onset of the pair-instability supernova mass gap}

\author{Ignacio Maga\~na~Hernandez}
\email{imhernan@andrew.cmu.edu}
\affiliation{\CMU}

\author{Antonella Palmese}
\email{apalmese@andrew.cmu.edu}
\affiliation{\CMU}

\begin{abstract}
We analyze binary black hole (BBH) mergers from the latest Gravitational Wave Transient Catalog (GWTC-3) using a flexible, non-parametric framework to infer the underlying black hole mass distribution. Our model employs Gaussian Processes (GPs) with astrophysically motivated priors to represent the mass distribution, assuming both component masses are independently drawn from a common mass function. This approach enables us to capture complex features in the population without relying on rigid parametric forms. Motivated by predictions from binary stellar evolution, we focus on the presence of a mass gap in the range $40-120~ M_\odot$, attributed to the pair-instability supernova (PISN) and pulsational PISN (PPISN) processes. Using our GP-based model, we find for the first time, strong evidence for the onset of this mass gap, locating its lower edge at approximately $45-60 ~M_\odot$. We observe a suppression in the BBH merger rate when comparing the latest constraints against our GP model, by a factor of $10-60$ at $\sim60~M_\odot$, corresponding to the minimum of the mass gap. Additionally, we find evidence of a subpopulation of mergers populating the mass gap around $70 ~M_\odot$, which we argue is due to hierarchical mergers, as well as of a feature in the $40-50~ M_\odot$ range, albeit with low significance, which we attribute to the predicted PPISN build up. We discuss the astrophysical implications of this result in light of GW231123, a recently reported BBH merger with a total mass potentially above the PISN gap, suggesting the need for revised models of massive stellar evolution or alternative formation channels.
\end{abstract}

\date{\today}
\keywords{gravitational-waves, populations}
\maketitle

\section{Introduction} \label{sec:Intro}

The direct detection of gravitational waves (GWs) from binary black hole mergers by the LIGO and Virgo collaborations \cite{LIGOScientific:2016vbw} has opened a new window into the study of compact object populations. With the release of the third LIGO/Virgo/KAGRA (LVK) Gravitational Wave Transient Catalog (GWTC-3; \cite{GWTC3}), the number of confidently detected BBH mergers has increased to nearly 90, enabling more precise inferences about the underlying distributions of black hole (BH) masses, spins, and redshifts~\citep{LIGOScientific:2020kqk,LIGOScientific:2021psn}.

One of the most elusive features predicted by stellar evolution is the presence of a mass gap in the black hole mass spectrum, resulting from pair-instability supernova (PISN) and pulsational pair-instability supernova (PPISN) processes ~\citep{Woosley:2002zz,Heger:2001cd,Heger:2002by,Woosley:2016hmi,2019ApJ...878...49W,Belczynski2016}. These mechanisms are expected to inhibit the formation of BHs within the approximate mass range of $40-120~ M_\odot$, depending on factors such as stellar metallicity and rotation. Observationally constraining the location and shape of this mass gap is critical for testing models of binary stellar evolution and for identifying potential alternative formation channels that could populate the gap—such as hierarchical mergers in dense stellar environments \cite{Gerosa2021}.

Recent detections have challenged the canonical picture of the PISN mass gap. In particular, the event GW231123~\cite{LIGOScientific:massgap}, reported in the latest observing run (O4), features a primary black hole with a mass potentially exceeding $60\,M_\odot$, placing it within or above the expected PISN gap~\cite{Croon:2025gol,Stegmann:2025cja,Cuceu:2025fzi,Delfavero:2025lup}. This observation calls into question the robustness of current population models and the assumptions underlying parametric fits—particularly the impact of the assumed functional form on the inferred mass distribution.

In this work, we introduce a flexible, non-parametric framework based on Gaussian Processes to infer the BBH mass distribution from GWTC-3. Our model assumes that both component masses are independently drawn from a common mass distribution, modeled as a GP with astrophysically motivated priors. This approach enables us to capture complex features in the mass spectrum—such as sharp cutoffs or localized bumps—without imposing rigid parametric forms. Previous works have used binned Gaussian processes to model the population of BBH mergers in GWTC-3~\cite{Ray:2023upk, ray2024searchingbinaryblackhole, MaganaHernandez:2024uty}, however, the large model complexity introduces large errors on the inferred population, making it difficult to probe more localized features.

We focus on identifying the onset of the PISN mass gap and present strong evidence for a suppression in the black hole mass distribution above $\sim 45-60 M_\odot$. This result differs from previous constraints~\cite{KAGRA:2021duu}, which do not find the onset of the PISN mass gap but locate a maximum BH mass around $\sim 80~ M_\odot$, challenging the theoretical predictions if it were to be interpreted as the lower edge of the PISN mass gap. We discuss the implications of this finding in the context of GW231123 and explore possible astrophysical interpretations, including the need for revised stellar evolution models or contributions beyond isolated binary formation channels.

This paper is organized as follows. In Sec.~\ref{sec:Methods}, we provide a brief overview of hierarchical population inference, introduce our GP-based framework, and describe the data used in our analysis. In Sec.~\ref{sec:Results}, we present our main results on the inferred mass distribution and the evidence for the PISN mass gap. In Sec.~\ref{sec:AstroImplications}, we discuss the astrophysical implications of our findings. Finally, in Sec.~\ref{sec:Conclusion}, we conclude with a summary and an outlook for future work.

\section{Methods} \label{sec:Methods}
\subsection{Hierarchical Population Inference }

To infer the population distribution of BBH mergers as a function of their primary mass ($m_1$), secondary mass ($m_2$), and redshift ($z$) in the source frame, we employ hierarchical Bayesian population inference. For comprehensive reviews of this framework, we refer the reader to \citep{Thrane_2019, Vitale_2021}, and provide a brief summary below.

The number density of BBH events can be written as,
\begin{equation}
     \frac{dN(m_1, m_2, z | \Lambda)}{dm_1 dm_2 dz} \propto \frac{dV_c}{dz}\bigg(\frac{T_\mathrm{obs}}{1+z}\bigg) \mathcal{R}_0(1+z)^\kappa p(m_1, m_2 | \Lambda)
\end{equation}
\noindent
where $\Lambda$ denotes the model hyperparameters describing the BBH mass distribution $p(m_1, m_2 | \Lambda)$. We assume that BBHs are distributed uniformly in differential comoving volume $dV_c/dz$ but evolve according to a power law proportional to $(1+z)^\kappa$ as in other works~\citep{Fishbach_2018,KAGRA:2021duu}. $T_\mathrm{obs}$ is the total observation time and the factor of $1/(1+z)$ converts source-frame time to detector-frame time. Finally, $\mathcal{R}_0$ is the local BBH merger rate at $z=0$, over which we marginalize following~\citep{Farr_2019,Mandel2019,LIGOScientific:2020kqk}.

We make use of the rate-marginalized posterior on the population hyperparameters \( \Lambda \), given by
\begin{align} \label{eqn:posterior}
\begin{split}
    p\left(\Lambda \mid \{d_i\} \right) &\propto \frac{p(\Lambda)}{\beta(\Lambda)^{N_\mathrm{obs}}} \prod_{i=1}^{N_\mathrm{obs}} \left[ \int \mathcal{L}\left(d_i \mid m_1, m_2, z \right) \right. \\
    &\left. \times \frac{dN(m_1, m_2, z \mid \Lambda)}{dm_1\,dm_2\,dz} \, dm_1\,dm_2\,dz \right],
\end{split}
\end{align}

\noindent
where \( \{d_i\} \) denotes the set of strain data from each of the \( N_\mathrm{obs} \) gravitational-wave events included in the analysis. The function \( \mathcal{L}(d_i \mid m_1, m_2, z) \) represents the likelihood of the data \( d_i \) given source-frame primary and secondary masses \( m_1, m_2 \) and redshift \( z \). The term \( \beta(\Lambda) \) accounts for selection effects by quantifying the fraction of sources detectable under the population model specified by \( \Lambda \).

The integral over the individual event likelihoods in Eq.~\ref{eqn:posterior} is approximated via importance sampling using $N_i$ single-event posterior samples from single event parameter estimation analyses with default priors $\pi(m_1, m_2, z) \propto d_L^2 (1+z)^2 \frac{dd_L}{dz}$ \cite{KAGRA:2021duu} where $d_L$ is the luminosity distance at the source redshift $z$. To estimate the detectable fraction, we follow~\cite{Tiwari_2018} and~\cite{Farr_2019}, and use the LVK sensitivity estimate injection campaign~\citep{KAGRA:2021duu}---we use importance sampling to compute $\beta(\Lambda)$ and make sure that it converges~\citep{Farr_2019}.

\subsection{Mass Model}
\label{sec:models}
Following~\cite{farah:2023swu}, we model the population distribution of primary and secondary black hole masses as a symmetric distribution \( p_{\rm{BH}}(m|\Lambda) \), combined with a pairing function \( f(m_1, m_2|\Lambda) \). Specifically, we consider the following family of models:
\begin{equation}
    p(m_1, m_2|\Lambda) \propto p_{\rm{BH}}(m_1|\Lambda)\,p_{\rm{BH}}(m_2|\Lambda)\,f(m_1, m_2|\Lambda),
\end{equation}
where the pairing function depends on the mass ratio \( q = m_2/m_1 \), and is given by:
\begin{equation}
    f(m_1, m_2|\beta) = \left( \frac{m_2}{m_1} \right)^\beta,
\end{equation}
with \( \beta \) representing the power-law slope of the mass ratio distribution.

For completeness, we define the primary and secondary mass marginalized posterior distributions as
\begin{equation}
    p(m_1|\Lambda) = \int p(m_1, m_2|\Lambda) \ dm_2
\end{equation}
and
\begin{equation}
    p(m_2|\Lambda) = \int p(m_1, m_2|\Lambda) \ dm_1.
\end{equation}

\subsection{Astrophysics-Informed GP Mass Model}
To implement the flexible modeling of \( p_{\rm{BH}}(m|\Lambda) \) introduced above, we adopt a non-parametric Gaussian Process (GP) framework using the \texttt{tinygp} library \cite{tinygp2022} with \texttt{dynesty} \cite{Speagle:2019ivv} for sampling. This approach allows us to capture complex features in the mass spectrum—such as gaps or excesses—without imposing restrictive parametric forms. We use a Kernel mixture to model the different features in the BH mass distribution. Mixtures of kernels provide enhanced flexibility in GP modeling by allowing the covariance structure to reflect multiple characteristic scales in the data \cite{Rasmussen2006}. In principle, an arbitrary number of kernel components could be used to model increasingly complex mass distributions. However, we find that a two-component mixture is sufficient to capture the features present in the current GWTC-3 dataset, including the onset of the PISN mass gap and the emergence of a high-mass subpopulation.

We start by defining our GP kernel as a sum of two components. Specifically, we use a linear combination of a radial basis function (RBF) kernel and a Matérn-5/2 kernel. Each kernel is characterized by its own amplitude \( A_i \) and length scale \( \ell_i \), enabling the model to simultaneously capture both broad and localized structures in the mass distribution.  Here, \( m \) and \( m' \) denote two points in the input space of black hole masses. The kernel function \( K(m, m') \) encodes the covariance between the population properties at these two mass values, reflecting how smoothly the underlying distribution varies across the mass spectrum. The kernel is given by
\begin{align}
    K(m, m^{\prime}|A_1,A_2,\ell_1,\ell_2) &= \nonumber
    A_1^2 K_{\rm{RBF}}(m, m^{\prime}| \ell_1)\\ &+ A_2^2  K_{\rm{Matern}}(m, m^{\prime}| \ell_2).
\end{align}

We place a Gaussian Process prior on the logarithm of the mass distribution, such that
\[
\log p_{\rm{BH}}(m) \sim \mathcal{GP}\left(\mu(m), K(m, m')\right),
\]
where \( \mu(m) \) is the mean function (which we set to zero-mean) and \( K(m, m') \) is the covariance kernel defined above. This choice allows for flexible, non-parametric modeling of the mass spectrum while preserving positivity of the inferred distribution. The GP is defined over a mass grid spanning \( [1, 150]\,M_\odot \).

To ensure physical plausibility—such as avoiding boundary effects common in GPs and enforcing well-measured minimum and maximum masses from GWTC-3 data—we apply smoothed low-mass and high-mass filters to the GP realization, following the form introduced in Eq.~B5 of~\cite{KAGRA:2021duu}. These filters gradually suppress the probability density below a minimum mass \( m_{\min} \) and above a maximum mass \( m_{\max} \), with smoothing scales \( \delta m_{\min} \) and \( \delta m_{\max} \), respectively. Additionally, we apply a fixed cutoff outside the range \( [2, 100]\,M_\odot \) to ensure consistency with the sensitivity estimates derived from the injection campaigns used for selection effect calculations. The resulting filtered GP realization allows for the construction of a normalized and valid probability density function \( p_{\rm{BH}}(m) \), which can then be interpolated on the mass grid and evaluated at the GWTC-3 mass samples for inference.

\section{Results} \label{sec:Results}

We apply our astrophysics-informed Gaussian Process (GP) model to the BBH events reported in GWTC-3, using posterior samples from the LVK public data release~\cite{gwtc3_data,gwtc-3pe}. Specifically, we use the 69 BBH detections with a false alarm rate (FAR) less than 1 per year, ensuring a high-confidence sample~\cite{KAGRA:2021duu}. 

\begin{figure*}
    \centering
    \includegraphics[width=\textwidth]{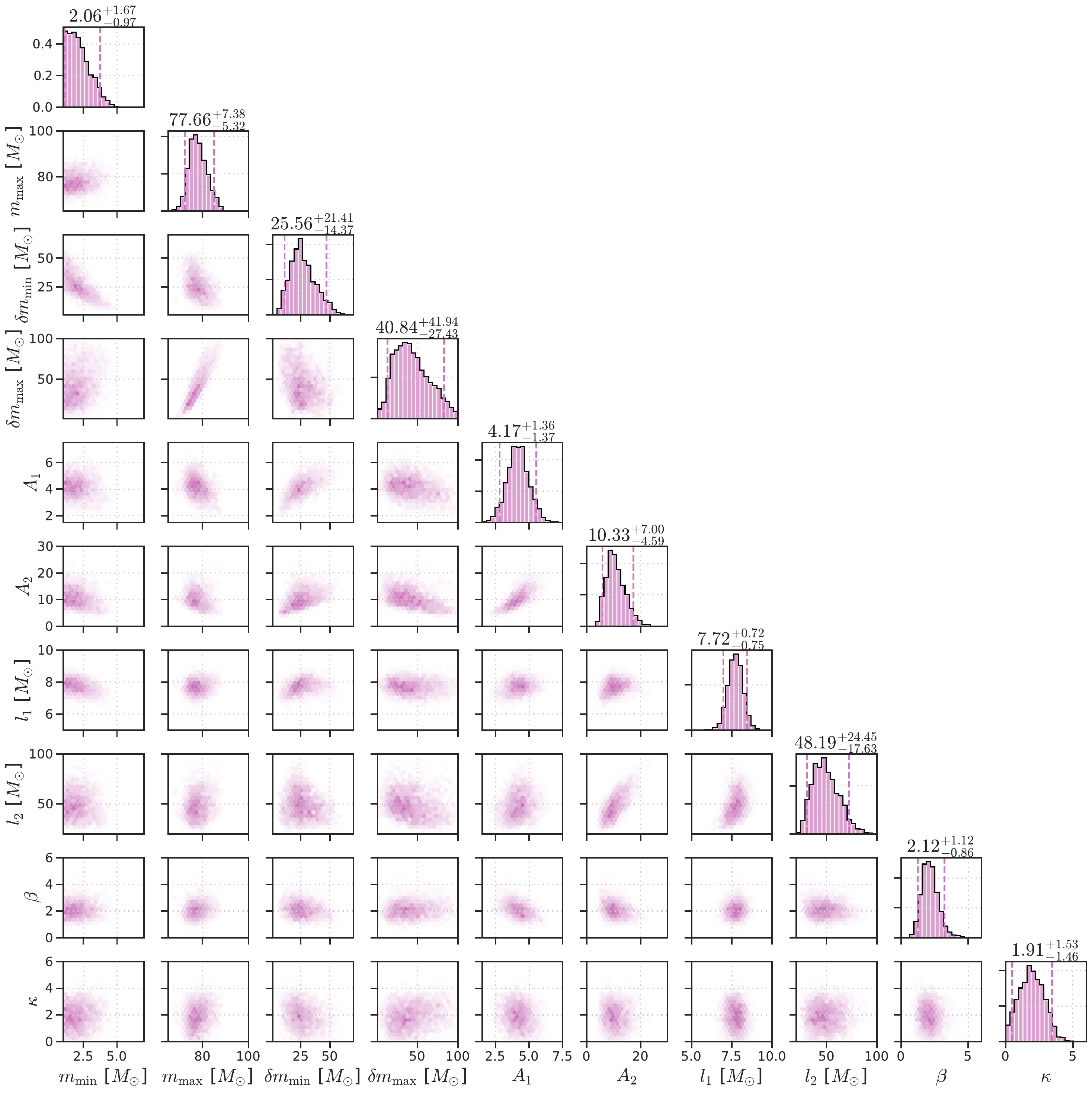}
    \caption{
        Corner plot showing the posterior distributions and covariances of the Gaussian Process model parameters used to infer the black hole mass spectrum. The parameters $A_1$ and $A_2$ represent the amplitudes of the RBF and Matérn-5/2 kernels, while $\ell_1$ and $\ell_2$ are their respective length scales, controlling the smoothness and locality of the inferred mass distribution. The parameters $m_{\mathrm{min}}$ and $m_{\mathrm{max}}$ define the low- and high-mass cutoffs, with smoothing widths $\delta m_{\mathrm{min}}$ and $\delta m_{\mathrm{max}}$. The power law slope for the mass ratio pairing function is given by $\beta$. The redshift evolution of the BBH merger rate is defined by the parameter $\kappa$. 
    }
    \label{fig:gp_corner}
\end{figure*}

In Figure~\ref{fig:gp_corner}, we show the posterior distributions for the Gaussian Process model parameters used to infer the BBH mass spectrum. The plot includes kernel amplitudes and length scales (\( A_1, A_2, \ell_1, \ell_2 \)), mass cutoffs and smoothing widths (\( m_{\mathrm{min}}, m_{\mathrm{max}}, \delta m_{\mathrm{min}}, \delta m_{\mathrm{max}} \)), the redshift evolution parameter \( \kappa \), and the pairing function power-law slope \( \beta \). The posteriors for \( \beta \) and \( \kappa \) are well-constrained and consistent with results from LVK and other studies~\cite{KAGRA:2021duu,Fishbach_2018,Fishbach:2019bbm,Callister:2023tgi,farah:2023swu}. We observe mild degeneracies among the kernel and filter parameters, reflecting the interplay between smoothness and boundary enforcement in the inferred mass distribution.

\begin{figure*}
\begin{center}
\includegraphics[width=\textwidth]{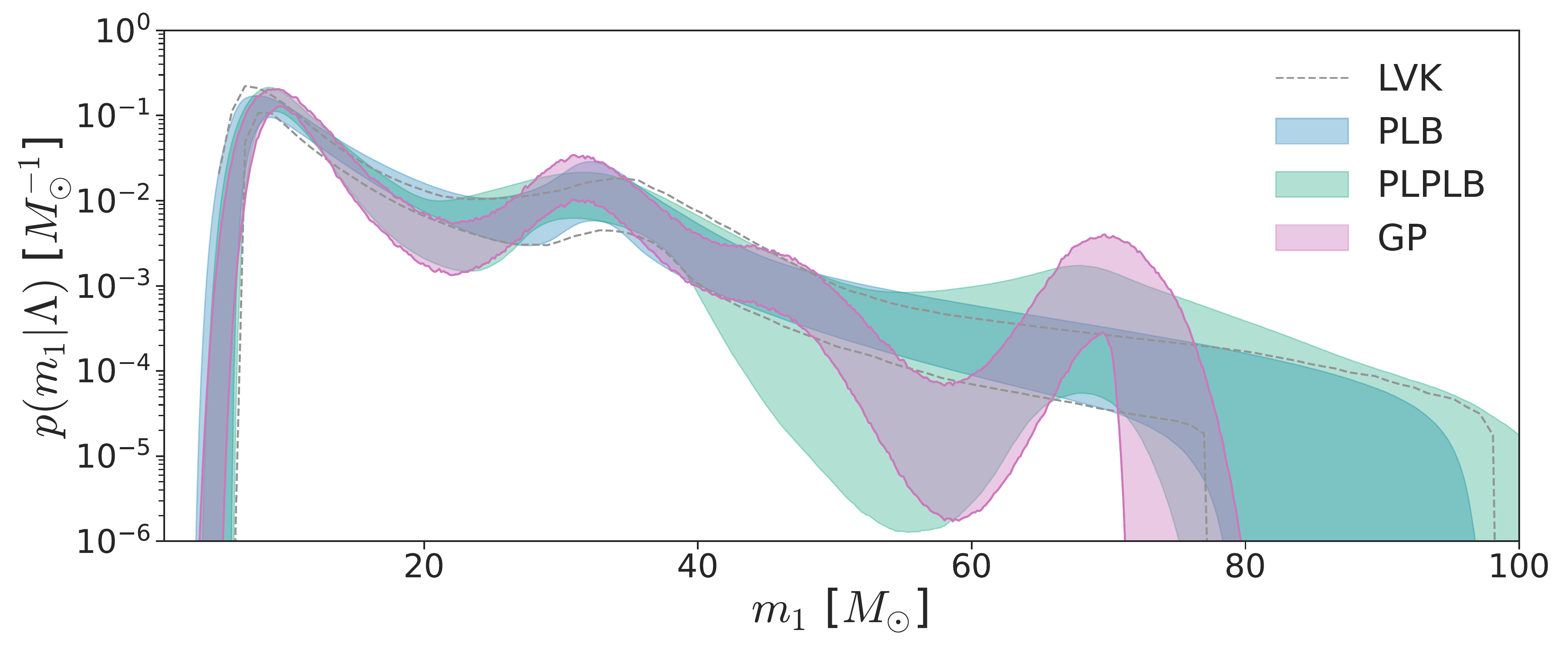}
\includegraphics[width=\textwidth]{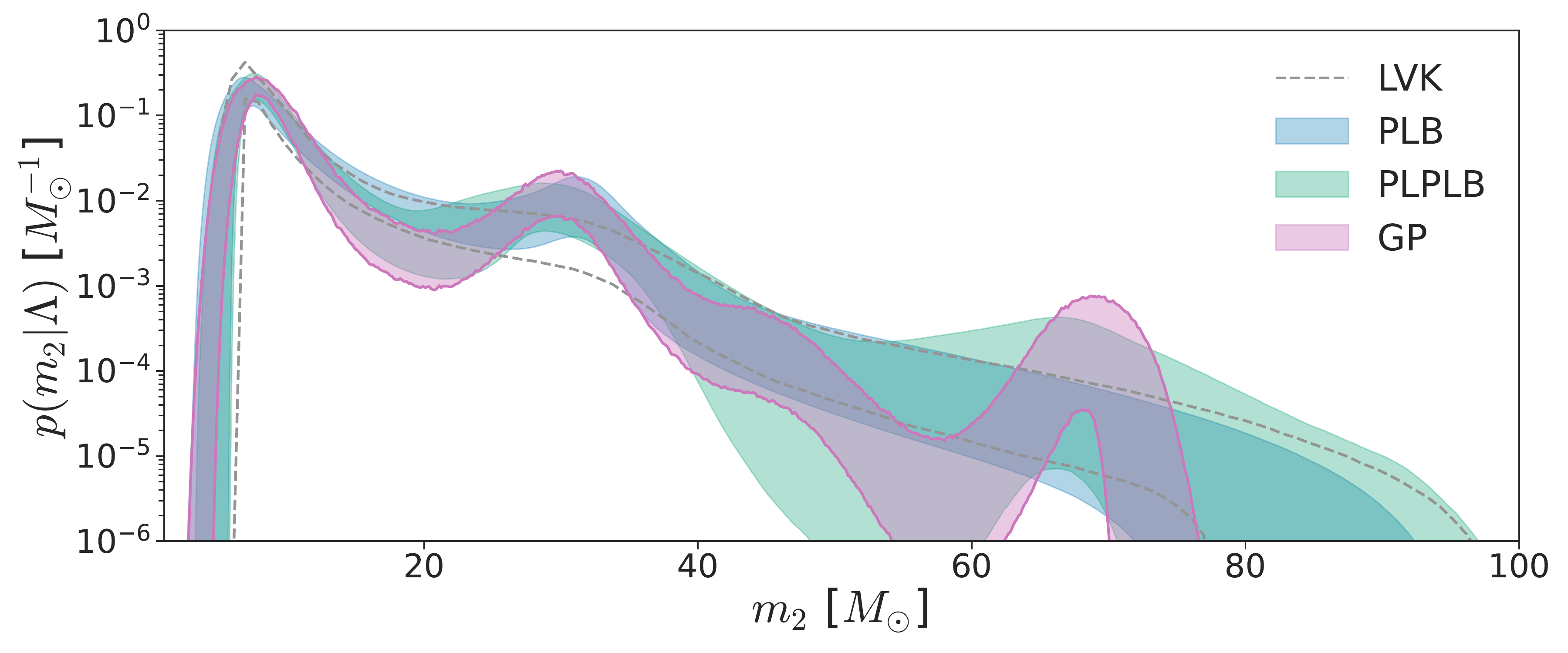}
\caption{\label{fig:gwtc3} Primary and secondary (marginalized) black hole mass distributions inferred from GWTC-3. The grey curves show the LVK population model posteriors, while the blue (PLB) and green (PLPLB) curves represent the flexible mixture model fits from ~\cite{MaganaPalmese2025}. The pink curve corresponds to our astrophysics-informed Gaussian Process model. Shaded regions indicate 95\% credible intervals. Our analysis reveals both a distinct subpopulation in the $60-80~ M_\odot$ range and strong evidence for the onset of the PISN mass gap between $45-60~ M_\odot$ in both component masses.}
\end{center}
\end{figure*}

\begin{figure}
\begin{center}
\includegraphics[width=0.46\textwidth]{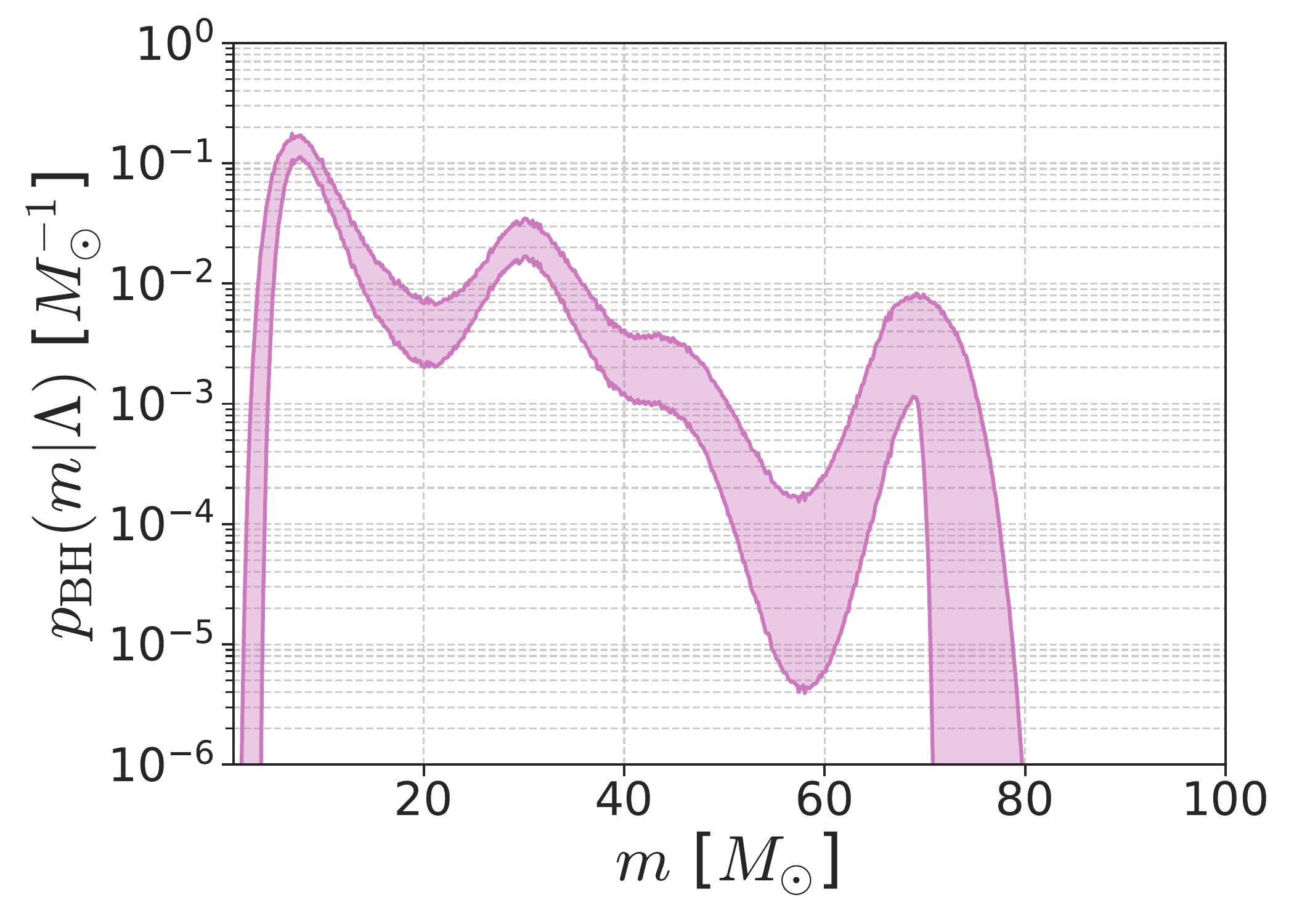}
\caption{\label{fig:massfunction} We show the posterior distribution on the common BH mass function $p_{\rm{BH}}$ inferred with GWTC-3 observations.}
\end{center}
\end{figure}

Figure~\ref{fig:gwtc3} shows the inferred primary and secondary mass distributions. The gray curves represent the LVK population model posteriors from GWTC-3 \cite{KAGRA:2021duu}, while the blue and green curves correspond to the \textsc{Powerlaw+Bump} (\plpeak) and \textsc{Powerlaw+Powerlaw+Bump} (\plplpeak) models—assuming a symmetric mass function as in this work—from~\cite{MaganaPalmese2025}. The purple curve shows the result from our GP-based model, with shaded regions indicating 90\% credible intervals. While our model recovers the general features of the LVK-inferred distribution, it reveals additional structure not captured by standard parametric forms. Additionally, in Figure \ref{fig:massfunction}, we show the common BH mass function, $p_{\rm{BH}}$ inferred from our analysis.

Most notably, we find strong evidence for a suppression in the mass distribution between the range $45-60~ M_\odot$, consistent with the onset of the PISN mass gap. This feature appears in both the primary and secondary mass distributions, owing to the symmetric mass model, and remains robust across different choices of priors on kernel hyperparameters. The inferred location of the lower edge of the gap is in slight disagreement with previous constraints which fail to capture the gap due to the use of parametric models. We can compare the 95\% confidence intervals at \( \sim 60\,M_\odot \) corresponding to the location within the gap with potentially highest depth. We find a suppression on the BBH merger rate by a factor in the range $10-60$ when comparing the \textsc{Powerlaw+Bump} model against our GP model. Additionally, we can compare the relative suppression with respect to the \( \sim 30\,M_\odot \) bump, present in all inferred populations. We find a suppression in the BBH rate by a factor in the range $400-3300$.

Additionally, our model identifies a distinct subpopulation of black holes in the $60-80~ M_\odot$ range, as reported in~\cite{MaganaPalmese2025}. This excess is not predicted by standard binary stellar evolution models and may point to alternative formation channels, such as hierarchical mergers in dense stellar environments \cite{Gerosa2021}. Comparing this feature at \( 70\,M_\odot \), we see that it is $5-20$ times higher in BBH merger rate compared to what the PLB model predicts. 

To assess the significance of these features, we compute the Bayes factor comparing our GP model to the PLB and PLPLB models from~\cite{MaganaPalmese2025}. We find \( \log_{10} \mathcal{B} = \{0.23, -0.06\} \) with estimated errors $\delta \log_{10} \mathcal{B} \approx 0.06$ for our GP model relative to the PLB and PLPLB models, respectively, indicating comparable support for our non-parametric approach. This supports the conclusion that the BBH mass spectrum may contain meaningful structure beyond what is captured by simple parametric forms. From a Bayesian perspective, our model offers sufficient complexity to capture both expected and unexpected features, while remaining competitive with commonly used parametric models in terms of model evidence and computational efficiency. We have checked that our results are consistent against our choice of priors on the GP hyperparameters and nested sampling settings. 

\section{Astrophysical Implications} \label{sec:AstroImplications}

Our analysis provides the first statistically significant evidence for the onset of the pair-instability supernova (PISN) mass gap in the binary black hole mass spectrum. We identify a sharp suppression in the distribution of both primary and secondary masses between $45-60~ M_\odot$, consistent with theoretical predictions for the lower edge of the PISN/PPISN mass gap \cite{Woosley2017, Belczynski2016, Marchant2019}. This result supports the long-standing hypothesis that stars with helium core masses in the range $40-120~ M_\odot$ undergo PISN/PPISN, leaving no black hole remnant \cite{Heger2002, Woosley2019}.

The detection of this mass gap has implications for our understanding of massive stellar evolution. It suggests that the physics of electron-positron pair production in stellar cores plays a dominant role in shaping the upper end of the black hole mass function \citep{LIGOScientific:2020ufj,Kimball_2020,Kimball:2020qyd}. The decline of the inferred mass gap onset suggests that the transition from successful collapse to complete disruption is modestly abrupt at most, in agreement with detailed stellar evolution models \cite{Farmer2019}.

However, the recent detection of GW231123 \cite{GW231123} might be challenging for this scenario. With component masses of $137^{+22}_{-17}\,M_\odot$ and $103^{+20}_{-52}\,M_\odot$, GW231123 has a primary mass that lies well within or above the PISN mass gap. However, the high spins of both black holes and the total remnant mass of $\sim 225\,M_\odot$ suggest a formation channel involving hierarchical mergers or dynamical assembly in dense stellar environments~\cite{GW231123}. Such scenarios can bypass the PISN limit by building up mass through successive mergers of lower-mass black holes \cite{Gerosa2021, Rodriguez2019}.

Our results also show a significant excess of black holes in the $60-70~ M_\odot$ range, a feature previously hinted in \cite{MaganaPalmese2025}. This bump may be due to a population of second-generation black holes formed via hierarchical mergers. These objects can contaminate the PISN mass gap, especially near its lower edge, making it difficult to isolate the stellar evolution population of BBH mergers. The presence of this subpopulation shows importance of flexible, non-parametric models—such as the Gaussian Process framework employed here—for disentangling different formation channels (at similar mass scales) in the BBH population.

It is worth mentioning that we also find some evidence for a potential feature in the range $40-50~ M_\odot$ albeit with much lower significance, due to the significant overlap of the 95\% confidence interval across all models. Under the interpretation that PISN processes lead to a mass gap in the predicted ranges, and given the results of our analysis, this feature can be interpreted as potentially driven by the PPISN process, which predicts a buildup of black holes forming near the lower edge of the PISN mass gap. The statistical significance of this feature is much lower than that of the observed mass gap and the $60-70~ M_\odot$ feature; however, analyses with upcoming gravitational wave observations may be able to probe whether this feature persists. 

While our results provide strong evidence for a suppression of the BBH rate consistent with PISN mass gap predictions, other astrophysical formation channels may contribute to populating this region with black holes that do not form via binary stellar evolution. Hierarchical mergers in dense stellar environments such as globular clusters, nuclear star clusters, or young massive clusters are one potential explanation. Black holes formed from previous mergers can undergo subsequent mergers, producing second- or third-generation black holes with masses above the PISN cutoff \cite{Rodriguez2019, Gerosa2021,palmese_conselice}. Another possibility is chemically homogeneous evolution (CHE) in rapidly rotating, low-metallicity binaries, which can prevent the development of a core-envelope structure and allow stars to collapse directly into more massive black holes \cite{Mandel2016, Marchant2016}. Although CHE can extend the upper mass limit of black holes, it typically does not produce objects in the full extent of the PISN gap unless metallicity is extremely low. Additionally, black holes embedded in active galactic nucleus (AGN) disks may grow via gas accretion or repeated mergers, enabling the formation of massive black holes in environments with high retention efficiency \cite{Yang2019,Ford:2021kcw}. Finally, primordial black holes (PBHs), formed from early-universe density fluctuations, are not subject to stellar evolution constraints and in principle can populate the gap \cite{Sasaki2016}, though current observational constraints limit their contribution to the overall BBH population.

The presence of a subpopulation in the $60-70~ M_\odot$ range~\cite{MaganaPalmese2025}, may reflect the imprint of these alternative formation channels, particularly hierarchical mergers, which can contaminate the PISN gap and complicate its interpretation as arising purely from binary stellar evolution.

\section{Conclusion and Future Prospects} \label{sec:Conclusion}

We have presented a non-parametric, astrophysics-informed Gaussian Process model to infer the binary black hole mass distribution using GWTC-3 observations. By considering a mixture of kernels, our model captures both global and local features in the mass spectrum, offering a flexible alternative to traditional parametric approaches. This framework enables us to reconstruct the underlying mass distribution while accounting for measurement uncertainties and selection effects using the well developed hierarchical Bayesian population inference framework.

Our analysis provides the first statistically significant evidence for the onset of the PISN mass gap, with a sharp suppression in the BBH mass distribution between $45-60~ M_\odot$. This is consistent with theoretical predictions from stellar evolution models that include pair-instability supernovae \cite{Woosley2017, Belczynski2016, Marchant2019, Farmer2019, Woosley2019}. The sharpness of the inferred gap supports the hypothesis that stars with helium core masses in the range $40-120~ M_\odot$ undergo violent mass ejection or complete disruption, leaving no black hole remnant, leading to a gap~\cite{Heger2002}.

In addition to the mass gap, we identify a statistically significant excess of black holes in the $60-70~ M_\odot$ range. This feature, previously hinted at in~\cite{MaganaPalmese2025}, is consistent with a subpopulation of second-generation black holes formed via hierarchical mergers in dense stellar environments \cite{Gerosa2021, Rodriguez2019}. Such mergers can contaminate the PISN mass gap, particularly near its lower edge, making it difficult to isolate the effects of stellar evolution alone \cite{Kimball_2020,Kimball:2020qyd}. The recent detection of GW231123 \cite{GW231123}, a BBH merger with component masses well likely above the upper edge of the PISN gap, further supports the presence of a dynamically assembled population capable of populating the gap.

Looking forward, the ongoing fourth observing run of the LIGO-Virgo-KAGRA collaboration is expected to significantly expand the catalog of BBH mergers\cite{LIGOScientific:2025slb}. Using the methods developed in this paper, we can test whether the onset of the PISN mass gap first reported here, stands an increasing number of BBH detection---or whether it becomes polluted from sources arising from the different formation channels we have discussed. We expect to see robust tests on the nature of the upper edge of the mass gap, as detections such as GW231123 become more common---however, this event already hints at the upper edge's existence. With more data we can further test and refine our models, however, with non-parametric models such as the GP model presented here---no detailed fine tuning is necessary as our predictions attempt to remain agnostic of the underlying (unknown) astrophysics---while folding in known or robust constraints.

\section{Acknowledgements}
The authors would like to thank Ariel Amsellem, Katie Breivik, Henry Chai, Chayan Chatterjee, Karan Jani, Anarya Ray and Soumendra Roy for useful comments. IMH is supported by a McWilliams postdoctoral fellowship at Carnegie Mellon University. This material is based upon work supported by the National Aeronautics and Space Administration under Grant No. 22-LPS22-0025. This research has made use of data or software obtained from the Gravitational Wave Open Science Center (gwosc.org), a service of the LIGO Scientific Collaboration, the Virgo Collaboration, and KAGRA. This material is based upon work supported by NSF's LIGO Laboratory which is a major facility fully funded by the National Science Foundation, as well as the Science and Technology Facilities Council (STFC) of the United Kingdom, the Max-Planck-Society (MPS), and the State of Niedersachsen/Germany for support of the construction of Advanced LIGO and construction and operation of the GEO600 detector. Additional support for Advanced LIGO was provided by the Australian Research Council. Virgo is funded, through the European Gravitational Observatory (EGO), by the French Centre National de Recherche Scientifique (CNRS), the Italian Istituto Nazionale di Fisica Nucleare (INFN) and the Dutch Nikhef, with contributions by institutions from Belgium, Germany, Greece, Hungary, Ireland, Japan, Monaco, Poland, Portugal, Spain. KAGRA is supported by Ministry of Education, Culture, Sports, Science and Technology (MEXT), Japan Society for the Promotion of Science (JSPS) in Japan; National Research Foundation (NRF) and Ministry of Science and ICT (MSIT) in Korea; Academia Sinica (AS) and National Science and Technology Council (NSTC) in Taiwan.

\bibliography{references}{}
\bibliographystyle{aasjournal}

\end{document}